\documentclass[published]{JHEP}
\JHEP{04(1999)031}
\newcommand{\be}{\begin{equation}}
\newcommand{\ee}{\end{equation}}
\newcommand{\ba}{\begin{eqnarray}}
\newcommand{\ea}{\end{eqnarray}}

\newcommand{\ga}{\dot{g}_A}
\newcommand{\text}{\mbox}
\newcommand{\NEG}{\not\!}

\title{Field redefinitions and wave function renormalization to
$O(p^4)$ in heavy baryon chiral perturbation theory}

\author{Joachim Kambor\thanks{Supported in part by Schweizerischer
				Nationalfonds and by the EEC-TMR
				Program, Contract No. CT98-0169.}\\
				
	Institut f\"ur Theoretische Physik, Universit\"at Z\"urich\\
	Winterthurerstr. 190, CH--8057 Z\"urich, Switzerland\\
	E-mail: \email{kambor@physik.unizh.ch}} 

\author{Martin Moj\v zi\v s\thanks{Supported by VEGA grant
	No.1/4301/97.}\\ 
	Department of Theoretical Physics, Comenius University\\
	Mlynsk\'a dolina, SK--84215 Bratislava, Slovakia\\ 
	E-mail: \email{mojzis@fmph.uniba.sk}}

\abstract{Mass- and wave-function renormalization is calculated to
order $p^4$ in heavy baryon chiral perturbation theory. Two different
schemes used in the literature are considered.  Several technical
issues like field redefinitions, non-transformation of sources as well
as subtleties related to the definition of the baryon propagator are
discussed. The nucleon axial-vector coupling constant $g_A$ is
calculated to order $p^4$ as an illustrative example.}
\received{February 17, 1999}
\accepted{April 28, 1999}

\begin{document}

\section{Introduction}

Heavy baryon chiral perturbation theory (HBChPT)~\cite{JM91,BKKM92}
allows for a systematic low energy expansion of one-nucleon Green
functions.  However, the matrix elements calculated in HBChPT are
frame dependent.  In order to obtain Lorentz invariant S-matrix
elements, the fully relativistic nucleon propagator has to be worked
out too.~\cite{EM97} So-called heavy nucleon sources cannot be
neglected but yield non-trivial contributions to the nucleon wave
function renormalization $Z_{\rm N}$ already at order $p^3$.

The frontier of HBChPT calculations presently lies at the order
$p^4$~\cite{p4calcs}, and in one exceptional case at order
$p^5$~\cite{McGB98}.  Further complete $p^4$ calculations are needed
in order to fully assess the convergence properties of the perturbative
series. The aim of the present paper is to provide the renormalized
parameters of the leading order chiral lagrangean, i.e.\ $m_{\rm N}$,
$Z_{\rm N}$ and $g_A$, to ${\cal O}(p^4)$ --- a prerequisite for any
such complete order $p^4$ calculations.

Two HBChPT lagrangeans widely used in the literature are considered.
These are the lagrangean given in~\cite{BKKM92,BKMrev95} (called BKKM
hereafter) and the form appearing in~\cite{EM96} (called EM). The
difference consists in the absence of equation of motion (EOM) terms
in EM, which have been eliminated by nucleon field redefinitions. The
EM-form has the advantage of containing less terms. At order $p^4$ the
reduction in complexity is already rather striking. Moreover, since EM
uses a minimal basis of counter terms, the number of independent
coupling constants can be inferred directly.  On the other hand, the
determination of the wave function renormalization constant $Z_{\rm
N}$ is more involved, in particular when going beyond order $p^3$. We
thus extend the work of ref.~\cite{EM97} and calculate mass and wave
function renormalization to order $p^4$, both for the BKKM as well as
the EM lagrangean. In the calculation of $Z_{\rm N}$ several
additional issues enter compared to the treatment in~\cite{EM97}.%
\footnote{Wave function renormalization to order $p^4$ in the BKKM
case was treated recently in~\cite{SMF98}. The emphasis in this
article is on different aspects than in the present work.}  The
related subtleties are exposed by introducing a new EOM-transformation
at the level of the relativistic lagrangean which allows for a direct
and elegant evaluation of $Z_{\rm N}$. We also comment on the
non-transformation of nucleon sources. Finally, we calculate the
nucleon axial-vector coupling constant $g_A$ to order $p^4$ in the two
schemes considered. Although different at intermediate steps, the
final results agree with each other, as expected. Some
phenomenological implications of this result are discussed.

\section{General formalism and EOM field transformations}

The starting point for the derivation of HBChPT is the generating
functional of relativistic Green functions
\begin{equation}
e^{i Z[j,\eta,\bar\eta]} = {\cal N} \int \left[du d\Psi d\bar\Psi\right] \exp\left\{ i 
\left(\tilde{S}_{\rm M}+ S_{\rm MB} +\int d^4x \left(\bar\eta \Psi+\bar\Psi \eta\right) \right)\right\} \,.
\label{genfunct}
\end{equation}
$j,\eta,\bar\eta$ denote the sources of mesonic and baryonic fields,
respectively. $\tilde{S}_{\rm M}$ is the mesonic action --- the tilde
reminds us of the nucleon degrees of freedom having not been
integrated out --- and $S_{\rm MB}$ is the action corresponding to the
pion nucleon lagrangean~\cite{GSS88}
\begin{equation}
{\cal L}_{\pi N} = \bar\Psi \left(i\not\!{\nabla}-m +{\ga \over 2} \not\!{u} 
\gamma_5\right) \Psi + \cdots\,,
\label{piNLag}
\end{equation}
where $m$ and $\ga$ denote the nucleon mass and axial-vector decay constant 
in the chiral limit, respectively. The ellipsis in~(\ref{piNLag}) stand for 
higher order terms.

A systematic low energy expansion is obtained by the frame dependent
decomposition of the nucleon field
\begin{equation}
\Psi(x) = e^{-i m v\cdot x} (N_v+H_v) (x)\,,
\end{equation}
with $v$ being a unit time-like four-vector and
\begin{equation}
P_v^+ N_v = N_v\,, \qquad P_v^- H_v = H_v\,, \qquad 
P_v^\pm = {1\over 2}(1\pm\not\!{v})\,.
\end{equation}
In terms of these fields the pion-nucleon effective action may be rewritten as
\begin{equation}
S_{\rm MB} = \int d^4x \left\{ \bar{N}_v A N_v + \bar{H}_v B N_v 
+\bar{N}_v \gamma_0 B^\dagger \gamma_0 H_v - \bar{H}_v C H_v\right \}\,.
\label{SMB}
\end{equation}
Introducing sources corresponding to $N_v,H_v$
\begin{equation}
\rho = e^{i m v\cdot x} P_v^+ \eta\,, \qquad R =  e^{i m v\cdot x} P_v^- \eta\,,
\end{equation}
the heavy components $H_v$ can be integrated out,~\cite{MRR92,BKKM92,E94}
yielding
\begin{eqnarray}
\label{intermediate}
e^{i Z[j,\eta,\bar{\eta}]} &=& {\cal N}' \int \left[du dN_v
d\bar{N}_v\right] \exp \Bigg\{ i \bigg(\tilde{S}_{\rm M} + \int d^4x
\Big[ \bar{N}_v \tilde{A} N_v + \bar{N}_v \gamma_0 B^\dagger \gamma_0
C^{-1} R +
\nonumber\\
& & \hphantom{{\cal N}' \int \left[du dN_v d\bar{N}_v\right] \exp
\Bigg\{} +\bar{R} C^{-1} B N_v +\bar{N}_v \rho+\bar{\rho} N_v +\bar{R}
C^{-1} R \Big]\bigg) \Bigg\}\,, \hphantom{(2.7)}
\end{eqnarray}
with 
\begin{equation}
\tilde{A} = A + \gamma_0 B^\dagger \gamma_0 C^{-1} B \,.
\label{Atildedef}
\end{equation}
Finally, expanding $C^{-1}$ in a power series in $1/m$ and integrating 
over $N_v$ yields the functional 
\begin{equation}
e^{i Z[j,\eta,\bar\eta]} = {\cal N}'' \int [du] e^{i\,
(S_{\rm M}+Z_{\rm MB}[u,j,\eta,\bar{\eta}])}\,,
\label{genfunctfin}
\end{equation}
where 
\begin{equation}
Z_{\rm MB}[u,j,\eta,\bar{\eta}] = -\int d^4x \Big[(\bar{\rho}+\bar{R} C^{-1} B)
\tilde{A}^{-1} (\rho+\gamma_0 B^\dagger \gamma_0 C^{-1} R) 
-\bar{R} C^{-1} R \Big]\,.
\label{ZMBfin}
\end{equation}

Any relativistic Green function is obtained from the functional
(\ref{genfunctfin}) by taking derivatives with respect to appropriate
sources.  The simplest example of this kind is the two-point function
of nucleon fields, which leads to a proper definition of the nucleon
mass and wave function renormalization. For more details we refer the
reader to ref.~\cite{EM97}.

The matrices A,B,C occurring in~(\ref{ZMBfin}) correspond to the
effective action~(\ref{SMB}). Explicit expressions for these have
appeared first in~\cite{BKKM92}, and we therefore call this set of
operators BKKM. An other form of the effective heavy baryon lagrangean
was introduced in~\cite{EM96}. It was shown that so-called equation of
motion terms in $\tilde A$ can be eliminated by a redefinition of the
``light component field'' $N_v$. However, in ref.~\cite{EM96} the
effect of such field redefinitions was studied on the level of the
effective lagrangean. The wave function renormalization, on the other
hand, depends also on the operators B and C. What is the effect the
EOM-transformations entail on B,C and hence on $Z_{\rm N}$?

In order to answer this question we re-investigate the nucleon field
redefinitions on the level of the generating functional.  We propose a
variant of the EOM-transformations employed in~\cite{EM96}, performed
on the relativistic nucleon fields $\Psi$. Since the formalism of
\cite{EM97} emphasizes the relativistic nature of the problem, this
seems to be more natural. We thus consider the field transformations
\begin{equation}
\Psi = (1 + e^{-i m v\cdot x} T e^{i m v\cdot x}) \Psi'\,,
\label{psitransform}
\end{equation}
with
\begin{equation}
T = P_v^+ T_{++} P_v^+ +  P_v^+ T_{+-} P_v^- +  P_v^- T_{-+} P_v^+ +  
P_v^- T_{--} P_v^-\,.
\label{+-decomp}
\end{equation}
The exponential factors in~(\ref{psitransform}) are introduced such
that possible derivatives in $T$ act on $N_v$ once the heavy baryon
variables are introduced. The decomposition~(\ref{+-decomp}) is useful
in order to separate the +/-- sectors of the theory.

The same steps as performed in Eqs.~(\ref{genfunct})--(\ref{ZMBfin})
can be carried through provided we set
\begin{equation}
T_{+-} = T_{-+} = 0\,.
\label{T+-choice}
\end{equation}
The generating functional then assumes the
form~(\ref{genfunctfin}),~(\ref{ZMBfin})\footnote{The Jacobian
associated with change of variables~(\ref{psitransform}) can be shown
to yield no contribution to S-matrix elements.}  but with the
replacements
\begin{eqnarray}
\label{ABCprime}
A \rightarrow A' &=& P_v^+ (1+\gamma_0 T_{++}^\dagger \gamma_0) P_v^+  A 
P_v^+ (1+T_{++}) P_v^+
\nonumber\\
B \rightarrow B' &=& P_v^- (1+\gamma_0 T_{--}^\dagger \gamma_0) P_v^-  B 
P_v^+ (1+T_{++}) P_v^+ 
\nonumber\\
C \rightarrow C' &=& P_v^- (1+\gamma_0 T_{--}^\dagger \gamma_0) P_v^-  C 
P_v^- (1+T_{--}) P_v^- \,,
\end{eqnarray}
as well as
\begin{eqnarray}
\label{rhoprime}
\rho \rightarrow \rho' &=& P_v^+ (1+\gamma_0 T_{++}^\dagger \gamma_0)
P_v^+ \rho \nonumber\\
R \rightarrow R' &=& P_v^- (1+\gamma_0 T_{--}^\dagger \gamma_0) 
P_v^- R \,.
\end{eqnarray}
Setting $T_{--}=0$ and choosing $T_{++}$ in accordance with the
explicit expression given in~\cite{EM96}, we recover the case where
the EOM-transformations are performed on the fields $N_v$, {\it
cf}~\cite{EM96}.  The EOM-terms in the effective heavy baryon
lagrangean $\tilde{A}'=\tilde{A}_{\rm EM}$ are then absent, by
construction. However, according to~(\ref{ABCprime}) the matrix
operator $B$ is also changed. Explicit calculations reveal that the
difference shows up first at ${\cal O}(p^4)$.

Consider now the EOM-transformations for general $T_{--}$. We still
have $\tilde{A}'=\tilde{A}_{\rm EM}$, i.e.\ the effective lagrangean
for the light component fields $N_v'$ is the same. The $T_{--}$ part
of the transformation~(\ref{psitransform}) therefore can be used to
bring the factors $C'^{-1} B'$ in~(\ref{ZMBfin}) to a convenient
form. The point here is that the last term in~(\ref{ZMBfin}),
$\bar{R}' C'^{-1} R'$, does not exhibit poles and thus yields no
contribution to S-matrix elements.

What is a convenient choice for $T_{--}$? In order to understand this question
we have to add one further element to the discussion. We choose $T_{--}$ such 
that the dressed nucleon propagator has standard form, i.e.\
\begin{equation}
S_{\rm N}(p) = { A(p^2) \not\!{p}+B(p^2) m_N \over p^2-m_N^2} \,.
\label{SNstandard}
\end{equation}
In general this need not to be true. In the present application, there
is an other four-vector at our disposal, namely $v_\mu$. The numerator
of~(\ref{SNstandard}) may then contain also a term of the form
$C(p^2)\not\!{v}$.  This actually happens if we use the
EOM-transformations~(\ref{psitransform}) with~(\ref{T+-choice}) and
$T_{--}=0$ --- the problem shows up first at order $p^4$. In this
situation, one has to find the eigenvectors and corresponding
eigenvalues of the dressed propagator in order to properly define the
appropriate factors for external legs.~\cite{tHoVe73} This problem can
be circumvented by exploiting the freedom to choose $T_{--}$, at least
to order $p^4$. The point is that only the field independent part of
the critical term $C^{-1} B$ is needed. Choosing
\begin{equation}
T_{--} = T_{++}
\label{T--choice}
\end{equation}
obviously yields
\begin{equation}
C'^{-1} B' = C^{-1} B  + 
{\rm field\ dependent\ terms}\,.
\end{equation}
Explicit calculations then show that the dressed propagator to ${\cal
O}(p^4)$ has standard form, hence the wave function renormalization
can be calculated as in~\cite{EM97} (cf.\ sect. 4). We conclude that
aside from the redefinition of sources, the
EOM-transformation~(\ref{psitransform}) together
with~(\ref{T+-choice}),~(\ref{T--choice}) leads to a generating
functio\-nal~(\ref{genfunctfin}) with effective $\pi N$ lagrangean as
constructed in~\cite{EM96} but is otherwise unchanged compared to the
BKKM case.

Finally we would like to comment on the significance of transformed
sources appearing in~(\ref{rhoprime}). The two-point function of
nucleon fields, for instance, is obtained by taking functional
derivatives of the generating functional with respect to the sources
$\eta$,$\bar{\eta}$. However, after applying the EOM-transformation
the generating functional is written in terms of transformed sources
$\rho'$, $R'$ or, equivalently, in terms of $\eta'$ with
\begin{eqnarray}
\eta' &=& \alpha\, \eta, \nonumber\\
\alpha &=& \left(1+e^{-i m v\cdot x} \left[P_v^+ \gamma_0 T_{++}^\dagger \gamma_0 P_v^+ +
                                 P_v^- \gamma_0 T_{--}^\dagger \gamma_0 P_v^-\right] 
e^{i m v \cdot x}\right)\,.
\label{alphadef}
\end{eqnarray}
The wave function renormalization as well as any Green function with two 
nucleon and arbitrary number of mesonic legs is therefore multiplied with
additional factors $~\alpha^\dagger \alpha$. When calculating S-matrix 
elements, these factors will be cancelled, however, because the Green function
has to be multiplied with two inverse nucleon propagators.%
\footnote{This intuitive argument can be put on a more rigorous footing by 
considering in detail the generating functional in the path integral 
formulation.}
The conclusion is that S-matrix elements are independent 
of the choice of sources. For practical purposes it is more convenient to 
use the transformed sources $\eta'$,$\bar{\eta}'$ for in 
this case the factor $\alpha$ in~(\ref{alphadef}) and its functional average
is not needed explicitly. We shall follow this prescription when calculating
$Z_{\rm N}$ to ${\cal O}(p^4)$ below.

\section{Effective lagrangean to O(p$^4$)}

Here we recollect all the terms of the effective $\pi $N-lagrangean
needed for the complete one-loop renormalization of the nucleon mass,
wave function and axial-vector coupling constant $g_A$.  The relevant
terms of the effective $\pi \pi$-lagrangean are well
known~\cite{GL84}.

The leading order relativistic $\pi $N-lagrangean was given
in~(\ref{piNLag}).  Higher order terms, corresponding to ellipses
in~(\ref{piNLag}), are
\begin{eqnarray}
{\cal L}_{\pi \rm N}^{(2)}&=&\overline{\Psi }\left\{ c_1\langle \chi
_{+}\rangle -\frac{c_2}{4m^2}\left( \left\langle u_\mu u_\nu \right\rangle
\nabla ^\mu \nabla ^\nu +\mbox{h.c.}\right) +\frac{c_3}2\left\langle u\cdot
u\right\rangle +\frac{c_4}4i\sigma ^{\mu \nu }\left[ u_\mu ,u_\nu \right]
\right\} \Psi 
\nonumber\\
{\cal L}_{\pi \rm N}^{(3)}&=&\frac b{(4\pi F)^2}\overline{\Psi }\frac
12\langle \chi _{+}\rangle \not\!{u}\gamma _5\Psi 
\nonumber\\
{\cal L}_{\pi
\rm N}^{(4)}&=&\frac d{m(4\pi F)^2}\overline{\Psi }\langle \chi _{+}\rangle ^2
\Psi\,, 
\end{eqnarray}
where we have displayed only those terms contributing to our
calculations. The LECs $c_i$ as well as $b$ and $d$ are
finite. Infinite parts of the 3rd and 4th order LECs (needed for
cancellation of loop infinities) are added explicitly in the heavy
baryon effective lagrangean. $b$ and $d$ are then renormalized LECs
with renormalization scale equal to the pion mass.

The HBChPT effective lagrangean in the BKKM version reads 
\begin{eqnarray}
\label{BKKM2}
\widehat{{\cal L}}_{{\rm BKKM}}^{(2)} & = & \bar N_v\Bigg\{ \frac
1{2m}(v\cdot \nabla )^2-\frac 1{2m}\nabla \cdot \nabla -
\frac{i\ga}{2m}\{S\cdot \nabla ,v\cdot u\}+c_1\langle \chi _{+}\rangle+
\\\nonumber  
 &  & \hphantom{\bar N_v\bigg\{}
+\left( c_2-\frac{\ga^2}{8m}\right) \left(
v\cdot u\right) ^2+c_3u\cdot u+\left( c_4+\frac {1}{4m}\right) 
i\varepsilon ^{\mu \nu \rho \sigma }u_\mu u_\nu v_\rho S_\sigma \Bigg\} N_v
\\
\label{BKKM3}
\widehat{{\cal L}}_{{\rm BKKM}}^{(3)} & = & \bar N_v\left\{ \frac i{4m^2}v\cdot
\nabla \left( (\nabla \cdot \nabla - v\cdot \nabla )^2 \right) +\frac b{(4\pi
F)^2}\langle \chi _{+}\rangle S\cdot u\right\} \;N_v
\\
\widehat{{\cal L}}_{{\rm BKKM}}^{(4)} & = & \bar N_v\Bigg\{ \frac
1{8m^3}(v\cdot \nabla )^2\left( \nabla \cdot \nabla - (v\cdot \nabla )^2
\right) +\\\nonumber 
&  &  \hphantom{\bar N_v\bigg\{}
 + \frac{c_1}{4m^3}\left( \overleftarrow{\nabla }_\mu \langle \chi
_{+}\rangle \nabla ^\mu -v\cdot \overleftarrow{\nabla }\langle \chi
_{+}\rangle v\cdot \nabla \right) +\frac d{m(4\pi F)^2}\langle \chi
_{+}\rangle ^2\Bigg\} N_v\,,
\end{eqnarray}
where only the finite part of the lagrangean was displayed. For infinite
terms see~\cite{E94} and~\cite{MMS98}.

As already mentioned, this effective lagrangean can be simplified
considerably by the EOM-transformations, leading to the EM version of the
HBChPT effective lagrangean. In the second order the simplification is only
modest, but in higher orders it is more impressive 
\begin{eqnarray}
\label{EM2}
\widehat{{\cal L}}_{{\rm EM}}^{(2)} & = & \bar N_v\frac 1m\Bigg\{ -\frac
12\nabla \cdot \nabla - 
\frac{i\ga}2\{S\cdot \nabla ,v\cdot u\}+a_1\left\langle u\cdot
u\right\rangle+  
\nonumber \\  &  &  \hphantom{\bar N_v\frac 1m\Bigg\{}
+a_2\left\langle \left( v\cdot
u\right) ^2\right\rangle +a_3\langle \chi _{+}\rangle +a_5i\varepsilon ^{\mu
\nu \rho \sigma }u_\mu u_\nu v_\rho S_\sigma \Bigg\} N_v 
\\
\label{EM34}
\widehat{{\cal L}}_{{\rm EM}}^{(3)}&=&\frac{\widehat{b}}{(4\pi F)^2}\bar
N_v\langle \chi _{+}\rangle S\cdot u\;N_v
\nonumber\\
\widehat{{\cal L}}_{%
{\rm EM}}^{(4)}&=&\frac{\widehat{d}}{m(4\pi F)^2}\bar N_v\langle \chi
_{+}\rangle ^2N_v\,. 
\end{eqnarray}
The relations between the LECs $a_i$ and $c_i$ are: $a_1=$ $\frac
12mc_3+\frac 1{16}\ga^2,$ $a_2=\frac 12mc_2-\frac 18\ga^2,$ $%
a_3=mc_1$ and $a_5=mc_4+\frac 14(1-\ga^2)$. The LECs $\widehat{b}$ and $%
\widehat{d}$ are divergent
\begin{eqnarray}
\widehat{b}&=&b-\frac{(4\pi F)^2}{m^2}\ga a_3+\left( \frac 12\ga +\ga^3\right)
\left[ \ln \frac{M }\mu +(4\pi )^2L(\mu )\right] 
\label{brelation}
\\
\widehat{d}&=&d-\frac{(4\pi F)^2}{2m^2}a_3^2-\frac 3{16}\left(
4a_1+a_2-4a_3+\frac 38 \ga^2\right) \left[ \ln \frac{M }\mu +(4\pi
)^2L(\mu )\right], 
\end{eqnarray}
where $M$ is the pion mass to leading order in the chiral expansion and
\begin{equation}
\label{L}L(\mu )=\frac{\mu ^{D-4}}{(4\pi )^2}\left\{ \frac 1{D-4}-\frac
12[\ln 4\pi +1+\Gamma ^{\prime }(1)]\right\}\,. 
\end{equation}

\section{Nucleon mass and wave-function renormalization}

Nucleon mass and wave-function renormalization are determined by
the nucleon self energy. In the EM framework the one-loop 
(fig.~\ref{self-energy}) and tree graph contributions are given by
\begin{equation}
\Sigma^{\rm EM} =\Sigma_{{\rm tree}}^{\rm EM}+\Sigma_{{\rm loop}}^{\rm EM}\,,
\end{equation}
with
\begin{eqnarray}
\label{S T EM}\Sigma _{{\rm tree}}^{{\rm EM}}&=&-\frac{k^2}{2m}-\frac{4M^2a_3}%
m-\frac{M^4\widehat{d}}{\pi ^2F^2m}
\\
\label{S L EM}
\Sigma _{{\rm loop}}^{{\rm EM}} & = & \frac{3 \ga^2}{4F^2}(D-1)\left\{
J_2(\omega )+\frac{M^2}{mD}\Delta +\frac 1{2m}\left[ k^2+\left(
1+8a_3\right) M^2-2\omega ^2\right] J_2^{\prime }(\omega )\right\}  -
\nonumber\\&&
-\frac{6M^2}{mF^2}\left( a_1+\frac 1Da_2-a_3\right) \Delta \,.
\end{eqnarray}
$k$ is the nucleon residual momentum defined by $p=m\cdot v+k$;  
$J_2(\omega )$, $\Delta$ are standard one-loop integrals explicitly given 
in e.g.~\cite{M98} and $\omega = v \cdot k$.

\FIGURE[b]{\small
\unitlength=0.25mm
\linewidth=0.4pt
\linethickness{0.4pt}
\begin{picture}(550.00,150.00)(-10.00,0.00)
\put(10.00,20.00){\line(0,1){100.00}}
\put(11.00,20.00){\line(0,1){100.00}}
\put(11.00,0.00){\makebox(0,0)[cc]{(a)}}
\put(10.00,70.00){\oval(40.00,40.00)[r]}
\put(110.00,20.00){\line(0,1){100.00}}
\put(111.00,20.00){\line(0,1){100.00}}
\put(111.00,0.00){\makebox(0,0)[cc]{(b)}}
\put(128.00,70.00){\circle{35.00}}
\put(210.00,20.00){\line(0,1){100.00}}
\put(211.00,20.00){\line(0,1){100.00}}
\put(211.00,0.00){\makebox(0,0)[cc]{(c)}}
\put(210.00,70.00){\oval(40.00,40.00)[r]}
\put(210.50,90.00){\circle*{7.00}}
\put(310.00,20.00){\line(0,1){100.00}}
\put(311.00,20.00){\line(0,1){100.00}}
\put(311.00,0.00){\makebox(0,0)[cc]{(d)}}
\put(310.00,70.00){\oval(40.00,40.00)[r]}
\put(310.50,70.00){\circle*{7.00}}
\put(410.00,20.00){\line(0,1){100.00}}
\put(411.00,20.00){\line(0,1){100.00}}
\put(411.00,0.00){\makebox(0,0)[cc]{(e)}}
\put(410.00,70.00){\oval(40.00,40.00)[r]}
\put(410.50,50.00){\circle*{7.00}}
\put(510.00,20.00){\line(0,1){100.00}}
\put(511.00,20.00){\line(0,1){100.00}}
\put(511.00,0.00){\makebox(0,0)[cc]{(f)}}
\put(528.00,70.00){\circle{35.00}}
\put(510.50,70.00){\circle*{7.00}}
\end{picture}%
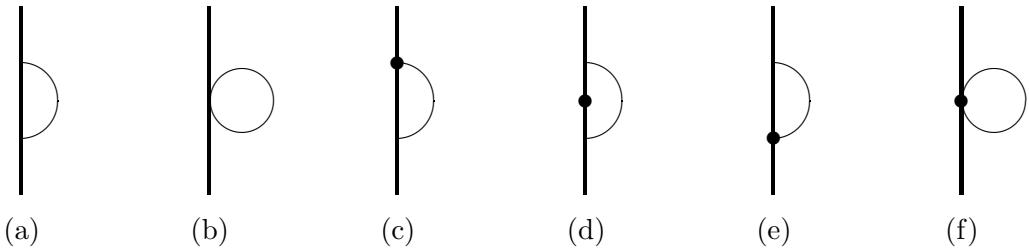
\caption{One-loop diagrams contributing to the nucleon self-energy.
Full circles are second order vertices. \label{self-energy}}}

To extract the nucleon mass and wave function renormalization, one has
to find the position and the residue of the pole of the nucleon
propagator.  Surprisingly, this procedure is not as straightforward as
one might expect and has become the subject of some discussions
recently~\cite{EM97,SMF98}.  The problem can be traced back to two
simple facts: there are two propagators, the relativistic and the
heavy baryon propagator. Moreover, the self-energy calculated in
HBCHPT is a function of two scalar variables, $\omega $ and $k^2$. The
relevant object to study is of course the relativistic propagator in
the variable $p^2$ --- HBChPT is, after all, just a particular way of
organizing the perturbation series of the full relativistic
theory. However, it requires some algebra to see that HBChPT
calculations lead to the relativistic propagator.

First, we rewrite the full HBChPT propagator as a function of the
variable $p^2$. To achieve this, we make use of the fact that the nucleon
self-energy contains (both in BKKM and EM versions) the term $-\frac
1{2m}(k^2+8M^2a_3)$. We can therefore write 
\begin{equation}
\frac i{\omega -\Sigma }\equiv \frac i{\Omega -\Sigma _{{\rm rest}}}\,, 
\end{equation}
where 
\begin{equation}
\Omega =\omega +\frac 1{2m}\left( k^2+8M^2a_3\right) =\frac 1{2m}\left(
p^2-m^2+8M^2a_3\right). 
\end{equation}
The next step is to trade $\omega $ for $\Omega $ also in $\Sigma
_{{\rm loop}}$. The crucial point here is that whenever a loop
integral $J_n\left( \omega \right) $ appears in the result, the
structure $\frac 1{2m}\left( k^2+8M^2a_3\right) \frac \partial
{\partial \omega }J_n\left( \omega \right) $ appears at higher order
as well (explicit results~(\ref{S L EM}) and~(\ref{S L BK}) provide
particular illustrations of this fact). The technical reason is that
insertion of the second order counter term into the nucleon propagator
inside the loop always contains $\frac 1{2m}\left(
k^2+8M^2a_3\right)$; the square of the propagator which enters due to
this insertion can then be written as the derivative of the propagator
with respect to $\omega$. Using the expansion
\begin{equation}
\Sigma _{{\rm loop}}\left( \Omega ,k^2\right) =\Sigma _{{\rm loop}}\left(
\omega ,k^2\right) +\frac 1{2m}\left( k^2+8M^2a_3\right) \frac \partial
{\partial \omega }\Sigma _{{\rm loop}}\left( \omega ,k^2\right) +\cdots 
\end{equation}
and the fact that the difference between $\omega $ and $\Omega $ is one
order higher than the omegas themselves,~(\ref{S L EM}) can be rewritten as 
\begin{eqnarray}
\Sigma _{{\rm loop}}^{{\rm EM}} & = & \frac{3\ga^2}{4F^2}(D-1)\left[
J_2\left( \Omega \right) +\frac 1{2m}\left( M^2-2\Omega ^2\right)
J_2^{\prime }\left( \Omega \right) +\frac 1D\frac{M^2}m\Delta \right] -
\nonumber\\  &  & 
-\frac{6M^2}{mF^2}\left( a_1+\frac 1Da_2-a_3\right) \Delta +{\cal O}\left(
p^5\right) \,.
\end{eqnarray}
Now one expands $\Sigma _{{\rm loop}}^{{\rm EM}}\left( \Omega \right) $
around the so-far unknown pole position $\Omega _p$ 
\begin{equation}
\omega -\Sigma ^{{\rm EM}}=\Omega +\frac{M^4\widehat{d}}{\pi ^2F^2m}-\Sigma
_{{\rm loop}}^{{\rm EM}}\left( \Omega _p\right) -\Sigma _{{\rm loop}%
}^{\prime {\rm EM}}\left( \Omega _p\right) \left( \Omega -\Omega _p\right)
+\cdots 
\end{equation}
This must vanish for $\Omega =\Omega _p$, i.e.\ 
\begin{equation}
\Omega _p=\Sigma _{{\rm loop}}^{{\rm EM}}\left( \Omega _p\right) -\frac{M^4%
\widehat{d}}{\pi ^2F^2m}\sim {\cal O}(p^3) \,.
\end{equation}
Consequently, $\Sigma _{{\rm loop}}^{{\rm EM}}(\Omega _p)=\Sigma _{{\rm loop}}^{{\rm EM}%
}\left( 0\right) +{\cal O}(p^5)$ and similarly for $\Sigma _{{\rm loop}%
}^{\prime {\rm EM}}(\Omega _p)$. One can therefore write
\begin{equation}
\label{HB propagator}\frac i{\omega -\Sigma ^{{\rm EM}}}=\frac{i2m\widehat{Z}%
}{p^2-m_N^2}+\cdots 
\end{equation}
with 
\begin{eqnarray}
m_N^2&=&m^2-8M^2a_3-\frac{2 M^4\widehat{d}}{\pi ^2 F^2}+ 2m \Sigma _{{\rm loop}%
}^{{\rm EM}}\left( 0\right) +{\cal O}(p^5)\,, 
\nonumber\\
\widehat{Z}&=&1+\Sigma _{{\rm loop}}^{\prime {\rm EM}}(0)+{\cal O}(p^4)\,.
\end{eqnarray}
More explicitly we have
\begin{eqnarray}
\label{m N}
m_N & = & m-\frac{4M^2a_3}m-\frac{M^3}{2(4\pi F)^2}\left[ 3\pi
\ga^2+\left( 32d-\frac 32a_2+\frac{21 \ga^2}{16}\right) \frac Mm\right] +
{\cal O}(p^5) \hphantom{(4.12)}
\\\label{Z hat}
\widehat{Z} & = & 1-\frac{3\ga^2 M^2}{2(4\pi F)^2}\left[ 1+3\left( \ln
\frac{M}\mu +(4\pi )^2L(\mu )\right) -\frac{3\pi }2\frac Mm\right] +
{\cal O}(p^4)\,.
\end{eqnarray}

Up to now we were dealing with the HBChPT nucleon propagator rather then the
full relativistic one, i.e.\ in the notation of~\cite{EM97} we have
considered only $S_{++}$. The full relativistic propagator is given by 
\begin{equation}
S_N = P_v^{+} S_{++} P_v^{+} + P_v^{+}S_{+-}P_v^{-} + 
P_v^{-} S_{-+} P_v^{+} + P_v^{-}S_{--}P_v^{-} \,.
\end{equation}
Proceeding along the lines of~\cite{EM97} one finds in the case of EOM
transformations defined by~(\ref{psitransform})--(\ref{T+-choice}) and~(\ref{T--choice}) 
\begin{eqnarray}
P_v^{+}S_{+-}P_v^{-} & = & P_v^{+}S_{++} 
\frac{\NEG k^{\perp }}{2m}\left( 1-\frac \omega {2m}+\frac{2M^2a_3}{m^2}%
\right) P_v^{-}+\cdots 
\nonumber \\ 
P_v^{-}S_{-+}P_v^{+} & = & P_v^{-}S_{++} 
\frac{\NEG k^{\perp }}{2m}\left( 1-\frac \omega {2m}+\frac{2M^2a_3}{m^2}%
\right) P_v^{+}+\cdots 
\nonumber\\ 
P_v^{-}S_{--}P_v^{-} & = & P_v^{-}S_{++}\frac{k^{\perp
}\cdot k^{\perp }}{4m^2}\left( 1-\frac \omega {2m}+\frac{2M^2a_3}{m^2}%
\right) ^2P_v^{-}+\cdots 
\end{eqnarray}
where ellipses stand for terms not contributing up to the fourth order
and $ \perp $ denotes perpendicular to $v,$ i.e.\ 
$X^{\perp}=X-v(v\cdot X)$.  Using the simple relation $X^{\perp }P_v^{\pm }=$
$P_v^{\mp }$ $X^{\perp }$ and the fact that $\omega $ is of the second
chiral order, one obtains
\begin{equation}
\label{SN}S_N=S_{++}\left\{ P_v^{+}+\frac{\not\!{k}^{\perp }}{2m}\left( 1-\frac
\omega {2m}+\frac{2M^2a_3}{m^2}\right) +P_v^{-}\frac{k^{\perp }\cdot k^{\perp }%
}{4m^2}+{\cal O}(p^4)\right\} .
\end{equation}

At this point we could continue as in~\cite{EM97} and write
$p=p_N+\lambda\, r$, where $p_N$ is the on-shell nucleon momentum and
$r$ is an arbitrary four-vector introduced to control the on-shell
limit.\footnote{We emphasize that in~\cite{EM97} a special choice
$r=v$ was used in the calculations, but one can check explicitly that
for~(\ref{SN}) the on-shell limit is independent of $r$.}  Here we
employ another method, which appears to be even simpler. We use yet
another decomposition of the nucleon momentum, $p=m_N\cdot v+Q$. This
implies $\not\!{p}+m_N=2m_NP_v^{+}+\not\!{Q}$. Moreover, on the mass
shell one has $2m_N v\cdot Q + Q^2 = 0$, i.e.\ $v \cdot Q = {\cal
O}(p^2)$, and therefore in the vicinity of the~pole
\begin{eqnarray}
S_N & = & S_{++}\!\left\{ \!
\frac{\NEG p+m_N}{2m_N}-\frac{\NEG v\;v\cdot Q}{2m_N}+\frac{\NEG Q}{2m_N}\!
\left( \frac{\delta m - v\cdot Q}{2m_N}+\frac{2M^2a_3}{m_N^2}%
\right)\! +P_v^{-}\frac{Q^2}{4m_N^2}+{\cal O}(p^4)\!\right\} 
\nonumber \\  
& = & 
S_{++}\left\{ \frac{\NEG p+m_N}{2m_N}-\frac{\NEG v\;v\cdot Q}{2m_N}-\frac{%
\NEG Q\;v\cdot Q}{4m_N^2}+\frac{Q^2}{8m_N^2}-\frac{\NEG v\;Q^2}{8m_N^2}+%
{\cal O}(p^4)\right\}, 
\end{eqnarray}
where $\delta m=m_N-m$. Collecting terms we finally have
\begin{equation}
S_N = S_{++} 
\frac{\NEG p+m_N}{2m_N}\left( 1+\frac{Q^2}{4m_N^2}\right) +{\cal O}(p^4) 
= \frac{\NEG p+m_N}{p^2-m_N^2} \widehat{Z}
\left(1-\frac{\delta m}{m_N}\right)\left(1+\frac{Q^2}{4m_N^2}\right)+\cdots 
\end{equation}
The dots correspond to higher orders and/or to terms vanishing at the 
pole.

We have arrived at the full relativistic nucleon propagator in the
form of the bare one, but with the bare mass replaced by the physical
one, and with an overall multiplicative factor, which is nothing else
but $Z_N^{{\rm EM}}$:
\begin{equation}
\label{Z EM}Z_N^{{\rm EM}}=\widehat{Z}-\frac{\delta m}{m_N}+\frac{Q^2}{%
4m_N^2}+{\cal O}(p^4)\,.
\end{equation}

In the BKKM framework the tree graph contribution to the nucleon self-energy~is
\begin{equation}
\label{S T BK}\Sigma _{{\rm tree,fin}}^{{\rm BKKM}}=-\frac 1{2m}\left(
k^2-\omega ^2\right) \left( 1-\frac \omega {2m}+\frac{\omega ^2}{4m^2}+\frac{%
4M^2c_1}{2m}\right) -4M^2c_1-\frac{dM^4}{m\pi ^2F^2}\,,
\end{equation}
where the subscript ``fin'' stands for the finite part of the tree
contribution and we refrain from giving explicitly the lengthy
expression for the infinite part. The loop graph contribution is,
after cancellations of extra terms coming from the difference between
$\widehat{\cal L}^{(2)}_{\rm BKKM}$ and $\widehat{\cal L}^{(2)}_{\rm
EM}$,
\begin{equation}
\label{S L BK}\Sigma _{{\rm loop}}^{{\rm BKKM}}=\Sigma _{{\rm loop}}^{{\rm EM}}\,.
\end{equation}

To proceed in analogy with the EM case one should rewrite $\Sigma
^{{\rm BKKM}}$ as a function of $\Omega $.  However, in $\Sigma_{{\rm
tree,fin}}^{{\rm BKKM}}$ it is impossible to get rid of $\omega$
completely. A simple trick to circumvent the problem is to replace
$1-\frac \omega {2m}+\frac{\omega ^2}{4m^2}$ by $(1+\frac \omega
{2m})^{-1}+{\cal O}(p^3)$, yielding
\begin{equation}
\label{S T BKa}
\omega -\Sigma _{{\rm tree,fin}}^{{\rm BKKM}}=
\left(1+\frac{\omega}{2m}\right)^{-1}\left[ \left(
1+\frac{4M^2c_1}{2m}\right) \Omega -\frac{M^2c_1}{ m^2 }
\Omega^2-\frac{16M^4c_1^2}{2m}+\frac{dM^4}{m\pi^2F^2}\right] .
\end{equation}

 From now on one can proceed as in the EM case. One obtains
again~(\ref{HB propagator}) with $m_N$ given by~(\ref{m N}) and with $%
\widehat{Z}$ replaced by
\begin{equation}
\widehat{Z}\rightarrow \left[\widehat{Z}+ \frac{9\ga^2 M^2}{2(4\pi
F)^2}\left(\ln\frac{M }\mu +(4\pi )^2L(\mu ) \right) \right] \left(
1+\frac \omega {2m}\right) \left( 1+\frac{\delta m}{2m}\right).
\label{ZhatBKKM}
\end{equation}
The terms in proportion to $\ln\frac{M }\mu$ and $L(\mu )$ in
(\ref{ZhatBKKM}) come from infinite EOM terms in the BKKM lagrangean
(not displayed explicitly in~(\ref{BKKM3})). The two multiplicative
factors on the RHS of~(\ref{ZhatBKKM}) cancel two terms in~(\ref{Z
EM}), which finally leads to
\begin{equation}
\label{Z BKKM}Z_N^{{\rm BKKM}}=
1-\frac{3\ga^2 M^2}{2(4\pi F)^2}+\frac{9\pi \ga^2 M^3 }{4 m(4\pi F)^2} +
{\cal O}(p^4)\,.
\end{equation}
This result agrees with the findings of~\cite{SMF98}.\footnote{The
result for $Z_N^{\rm BKKM}$ seems to be at variance
with~\cite{EM97}. However, as explained in the Erratum~\cite{EM97},
the third order BKKM lagrangean in~\cite{EM97} is not equivalent to
our eq.~(\ref{BKKM3}).}

\section{The nucleon axial-vector coupling constant $g_A$ to ${\cal O}(p^4)$ }

So far we have calculated nucleon wave-function and mass
renormalization up to the 4th chiral order. Renormalization of these
parameters of the leading order chiral lagrangean is to be used in any
complete one-loop HBChPT calculation.  On the other hand, the leading
order chiral lagrangean contains yet another parameter, namely the
nucleon axial-vector coupling constant $g_A$, which will also enter
any complete one-loop result. It is therefore equally worth to
calculate the relation between bare and physical $g_A$ up to the 4th
order.

$g_A$ receives contributions from both, tree and one-loop 
graphs. Moreover, at the order we are working, the wave function 
renormalization enters too. Working out $g_A$ for the two lagrangeans 
considered thus provides a consistency check on our results for $Z_{\rm N}$.

In the heavy baryon formalism, the matrix element of the iso-vector 
axial-vector current is given by~\cite{EM97}
\begin{eqnarray}
\Big\langle p_{\rm out}|\bar{q} \gamma^\mu \gamma_5 \tau^a q|p_{\rm in}\Big\rangle
\!&=&\!
\left( 1-{t\over 4 m_N^2}\right)^{-1} \bar{u}_+(p_{\rm out})\tau^a \times
\\\nonumber
&&\!\times\!\left\{\left[ 2\! \left( \!1-{t\over 4 m_N^2}\right) \!S^\mu \!- 
{q\cdot S \over m_N} v^\mu \right]\! G_A(t)\!+
{q\cdot S \over 2 m_N^2} q^\mu G_P(t) \right\} u_+(p_{\rm in}).
\end{eqnarray}
Concentrating on $g_A=G_A(0)$ we put $t=0$. Furthermore, we need only
that part of the the form factor in proportion to $S_\mu$. In
particular, pion-pole diagrams are $\sim q^\mu$ and need not to be
considered.

The relevant one-loop diagrams are those of fig.~\ref{self-energy}
with axial source hooked on in all possible places. Since the
lagrangeans of EM and BKKM are different, individual diagrams will in
general yield different results.  We obtain for the sum of all
one-loop graphs
\begin{eqnarray}
g_A^{\rm loop,\,EM} =&& {M^2 \over F^2} \left\{ 
{\ga \over 2} \left( \ga^2-4 \right) 
\left[ L+{1\over 16 \pi^2}\ln {M\over\mu} \right] 
+{\ga^3 \over 32 \pi^2}\right\}+ \nonumber\\
&& +{M^3 \over m F^2 \pi} \left\{ -{\ga^3\over 192} +{\ga \over 3}
\left( -a_1+a_5+{1\over 8}\right) \right\} \,.
\end{eqnarray}

In the BKKM case, there are additional loop-contributions due to the
EOM-terms in $\widehat{\cal L}^{(2)}_{\rm BKKM}$, {\it cf}
Eq.~(\ref{BKKM2}).  Moreover, this lagrangean is written in terms of
coupling constants $c_i$. This yields the difference
\begin{equation}
g_A^{\rm loop,\,BKKM}-g_A^{\rm loop,\,EM} = {M^3 \ga^3 \over m F^2 \pi}
\,{3\over 32}\,.
\label{loopdiff}
\end{equation}

The tree graph contribution is obtained from~(\ref{BKKM3}),(\ref{EM34}),
(\ref{brelation}) and~\cite{E94}, yielding
\begin{equation}
g_A^{\rm tree,\,EM} = \ga +{\widehat{b} \over (4\pi F)^2}\,4 M^2
\end{equation}
and 
\begin{equation}
g_A^{\rm tree,\,BKKM}-g_A^{\rm tree,\,EM} = {4 M^2 \over m^2}\,\ga a_3
-\frac{9\ga^3 M^2}{2(4\pi F)^2}\left( \ln\frac{M }\mu +(4\pi )^2L(\mu )
\right) .
\label{treediff}
\end{equation}
Applying wave function renormalization finally yields a third piece
\begin{equation}
g_A^{\rm Z_N,\,EM} = \ga\,\left( Z_N^{\rm EM}(0)-1\right),
\end{equation}
with $Z_N^{\rm EM}$ given in~(\ref{Z EM}). The difference between the two 
schemes here reads
\begin{equation}
g_A^{\rm Z_N,\,BKKM}-g_A^{\rm Z_N,\,EM} = -{4 M^2 \over m^2}\,\ga a_3 
-{M^3 \ga^3 \over m F^2 \pi}\,{3\over 32}
+\frac{9\ga^3 M^2}{2(4\pi F)^2}\left( \ln\frac{M }\mu +(4\pi )^2L(\mu )
\right).
\label{ZNdiff}
\end{equation}
We observe that the differences in
eqs.~(\ref{loopdiff}),(\ref{treediff}) and~(\ref{ZNdiff}) exactly cancel. 
Renormalizing the ${\cal O}(p^3)$ coupling constant $\widehat{b}$ according
to~(\ref{brelation}) we find in both schemes
\begin{equation}
g_A=\ga + \frac{M^2}{(4 \pi F)^2}(4 b -\ga^3)
+ \frac{M^3}{m(4 \pi F)^2}\frac{2\pi}{3} 
\left[ \ga (1 - 8 a_1 + 8 a_5) + \frac{11}{2} \ga^3 \right] . 
\label{g_A}
\end{equation}
Although different at intermediate steps, the final results for $g_A$ agree
with each other. This is the consistency check announced above.

The result~(\ref{g_A}) has important phenomenological consequences.
While the constant $b$ in the second term on the right hand side is
presently unknown, a rather precise estimate for the term in
proportion to $M^3$ can be given. The counter term coupling constants
$a_1$,$a_5$ are constrained from the nucleon sigma term and $\pi
N$-scattering threshold parameters \cite{M98,FMS98},
\begin{eqnarray}
a_1 &=& -2.6 \pm 0.7 \nonumber\\
a_5 &=&  3.3 \pm 0.8 \,,
\end{eqnarray}
where error bars significantly larger than in~\cite{M98,FMS98} have
been assigned. The reason for these larger error bars is
twofold. First, the well known problem with unrealistic error bars of
$\pi N$-scattering threshold parameters reflects itself in too
optimistic error bars of $a_1$,$a_5$. Second, the values of
$a_1$,$a_5$ were determined within the 3rd chiral order calculation
and their higher order corrections are presently unknown. In order to
account for both of these uncertainities, we take for the error bars a
conservative estimate of 25 \%.  Expanding consistently to higher
orders and employing input parameters $g_A=1.26$, $M=0.14$ GeV,
$m=0.939$ GeV and $F=0.093$ GeV we obtain a positive large correction
\begin{equation}
\Delta g_A |_{M^3} = 0.32 \pm 0.05\,,
\end{equation}
where the error bar is dominated by the assumed uncertainty in
$a_1$,$a_5$.  This surprisingly large correction implies that order
$p^3$ calculations carry potentially large uncertainties due to the
finite renormalization of $\ga$ entering at next order in the chiral
expansion.

\section{Conclusions}

We have calculated nucleon mass-, wave function and axial-vector
coupling constant renormalization to ${\cal O}(p^4)$ in heavy baryon
chiral perturbation theory. Relations~(\ref{m N}),~(\ref{Z
EM})/(\ref{Z BKKM})  
and~(\ref{g_A}) between bare and correspondig
physical values of these quantities are to be used in any complete
one-loop HBChPT calculation.

Two lagrangeans widely used in the literature have been considered.
These are the lagrangean given in~\cite{BKKM92,BKMrev95} and the form 
appearing in~\cite{EM96}. For the former case, we have confirmed the already
known results for mass- and wave function renormalization. New results were
obtained in the latter case, where the lagrangean is simpler because so-called
EOM-terms have been eliminated by nucleon field redefinitions. In this case,
however, wave function renormalization is more involved due to subtleties 
arising in conjunction with the field redefinitions. 
We have proposed a new EOM-field transformation,
performed on the relativistic nucleon fields. This yields the effective 
heavy baryon lagrangean given in~\cite{EM96} but also allows for a simple and 
elegant evaluation of $Z_{\rm N}$. 
Our result~(\ref{Z EM}) for $Z_{\rm N}$ associated to the lagrangean given 
in~\cite{EM96} enables a systematic use of this simpler version of 
lagrangean in future calculations.

We have completed renormalization of the parameters of the leading order chiral
lagrangean by calculating the nucleon axial-vector coupling constant $g_A$ to 
order $p^4$ in the two schemes considered. In this way, we have also checked 
our previous results --- although different at intermediate steps, the final 
result for $g_A$ is the same. Phenomenologically, the order $M_\pi^3$ 
correction to $g_A$ turns out to be rather large, $\Delta g_A |_{M_\pi^3} 
\simeq 0.32$. This might have important phenomenological consequences when 
going beyond order $p^3$ in HBChPT. 

\acknowledgments
We would like to thank G.~Ecker and S.~Steininger for clarifying discussions 
and T.R.~Hemmert for the collaboration at an early stage of this work.

\end{document}